\documentclass[pdflatex,12 pt]{article}
\usepackage{microtype}
\usepackage{subfigure}
\usepackage{graphicx}
\usepackage{epsfig}
\usepackage{epstopdf}
\usepackage{amssymb}
\usepackage{amsmath}
\usepackage{lipsum}
\RequirePackage[colorlinks,citecolor=blue,urlcolor=blue,linkcolor=blue]{hyperref}
\usepackage{dcolumn}
\usepackage{fullpage}
\title{\textbf{First principles correspondence of a second type of phonon anomaly along [211] to the Fermi nesting features and associated electron-phonon interactions in Ni$_2$FeGa}}
\author{Satyananda Chabungbam, Munima B. Sahariah\\
\textit{Institute of Advanced Study in Science and Technology}\\ Guwahati-35, INDIA}
\date{}

\begin{document} 

\maketitle

\begin{abstract}
First principles calculation supports the occurence of a phonon anomaly along [211] in Ni$_2$FeGa that was observed experimentally for the first time as structural modulations along [211] direction. Fermi surface scans have been performed in both austenite and martensite phase to observe the possible Fermi nesting features in the system. The magnitude of observed Fermi surface nesting vectors in (211) plane exactly match the phonon anomaly wavevectors along [211] direction. Electron-phonon calculations in the austenite phase shows that there is significant electron-phonon coupling in the system which may arise out of the lattice coupling between lower acoustic phonon modes and higher optical phonon modes and also because of strong Fermi surface nesting  features observed in the system.
\end{abstract}
\setlength{\columnsep}{0.8cm}
\section{INTRODUCTION}
Since the experimental and technological realization of magnetic shape memory alloys(SMA) based on Ni$_2$MnGa alloy of Heusler type in the last few decades, these class of materials have become a center for attraction to the world of metals and metallurgy as it has shown many novel properties like the shape memory effect, magnetocaloric effect, magnetoelastic effect etc \cite{Buc,Gho,Hu,Uij,Cla}. Ni$_2$FeGa has been considered a better alternative to Ni$_2$MnGa as SMA  with respect to ductility, higher transition temperatures, low transformation stress, high reversible strain, and small hysteresis\cite{Liu,Ham,Pal,Seh,Oik}. These alloys undergo a structural diffusionless phase transformation from austenite phase to martensite phase. Twin boundaries are developed in the martensite phase and the mobility of these martensitic domains leads to remarkably large macroscopic deformations\cite{Far}. So material scientists are now trying to explore the microscopic parameters that drive this technologically important martensitic transformation. Previous investigations in these alloys show that there is a metastable phase known as premartensite phase that exists in the form of modulated superstructures\cite{Oik,Oik2}. The presence of these phases can also be related to the twin boundaries that exist in the pre-martensite phase. These modulated supestructures are made up of layers of parallel monoclinic planes and accordingly, it may be 3M, 5M, 6M, 7M type of modulations. The superstructures in these type of systems are generally associated with electronic or lattice anomalies or sometimes both. The microscopic factors which can be related with these type of superstructures are the phonon anomaly in some particular branch of phonon spectrum, the Fermi surface nesting vectors, the susceptibility peaks, electron phonon coupling, etc. The premartensite phase is very rich in micromodulated structures and shows many competing phases. Studies using Transmission Electron Microscope shows that two kinds of micromodulated structures exists in Ni$_2$FeGa, one with Ga-rich and other with Fe-rich\cite{Hua}.  Electron diffraction experiment\cite{Li} shows diffused satellite peaks that correspond to micromodulated structures along both [110] and [211] directions in Ni$_2$FeGa Heusler alloy. The first one refers to the well known phonon anomaly along [110] TA$_2$ branch but the second is reported for the first time and it is believed to be associated with the phonon anomalies in the [211] branch. In an attempt to explore the above reported phonon anomaly along [211] direction, a thorough scan of the various possible phonon branches in both austenite and martensite phases of Ni$_2$FeGa system has been done. Fermi surface has been calculated for both upspin and downspin bands in both the phases and plotted in 3D and 2D to find out any possible Fermi nesting vectors. Electron-phonon coupling constant has been calculated in the austenite phase to check whether there is significant interaction of the electrons with the lattice or not. 
\section{COMPUTATIONAL DETAILS}
In this work, we use DFT formalism \cite{ph,wk} to calculate the spin polarised electronic structure as well as the dynamical properties. All the calculations are done using the Quantum Espresso package \cite{pg,ss}. The Perdew-Burke-Ernzerhof (PBE) functionals\cite{jp} are used to address exchange-correlation interactions. Ultrasoft pseudopotentials are taken for Ni( 3d$^8$4s$^2$), Fe( 3d$^7$4s$^1$), and Ga( 4s$^2$4p$^1$) to account the valence-core electrons interactions. For total energy calculation, the optimized plane wave kinetic energy cutoff is fixed at 40 Ryd and the density cutoff for ultrasoft pseudopotentials is fixed at 480 Ryd. The Methfessel-Paxton smearing technique for Brillouin zone integration has been employed for this metallic system. All structural and electronic parameters are well converged over a k-mesh value of $12\times12\times12$ Monkhorst-Pack grid. For vibrational studies, the density functional perturbation theory (DFPT) implemented in Quantum Espresso is used to calculate the frequencies. A well converged lattice dynamics calculation has been performed with a phonon convergence cut-off of $10^{-18}$ Rydberg. A q-mesh of $6\times6\times6$ generating 18 q-points have been used for phonon calculations in the austenite phase. For the martensite phase, a q-mesh of $4\times4\times4$ generating 16 q-points have been used. The interatomic force constants are calculated by Fourier transformation of the dynamical matrices generated at each of the \textit{q} points and finally interpolated back to get the full phonon dispersion spectra. For Fermi surface calculation, a uniform grid of $12\times12\times12$ is used for scf calculation for both phases and a higher grid of $26\times26\times26$ for austenite phase and  $24\times24\times24$ for martensite phase are used for nscf calculation. XCrySDen\cite{ak} visualization software is used to find the bands crossing the Fermi level.  Eliasberg's function has been calculated to compute the electron-phonon coupling constant in the austenite phase.


\section{RESULTS AND DISCUSSION}
\begin{figure}[ht]
  \centering
  \subfigure[L2$_1$]{\includegraphics[scale=0.159]{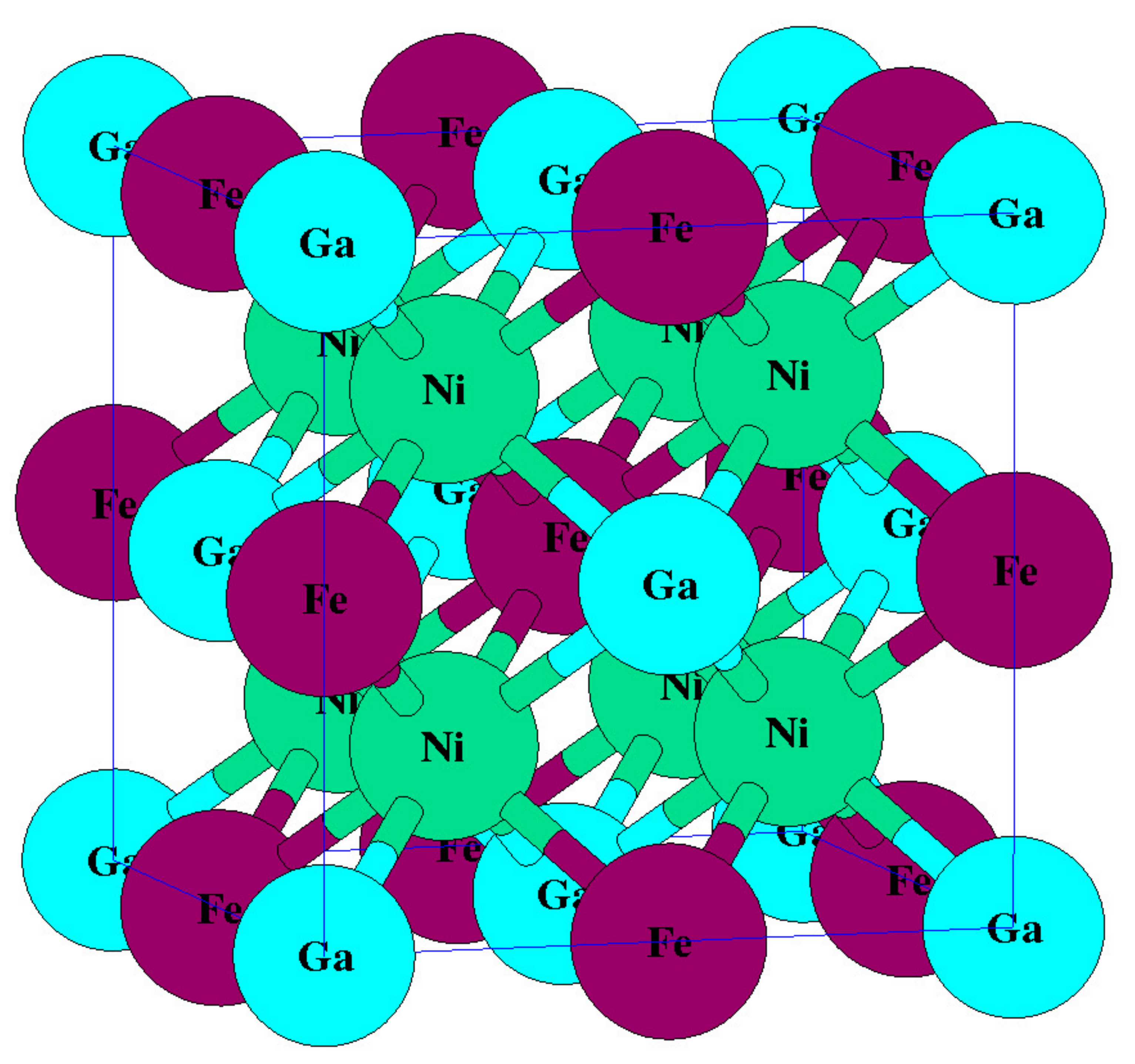}}
   \subfigure[L1$_0$]{\includegraphics[scale=0.143]{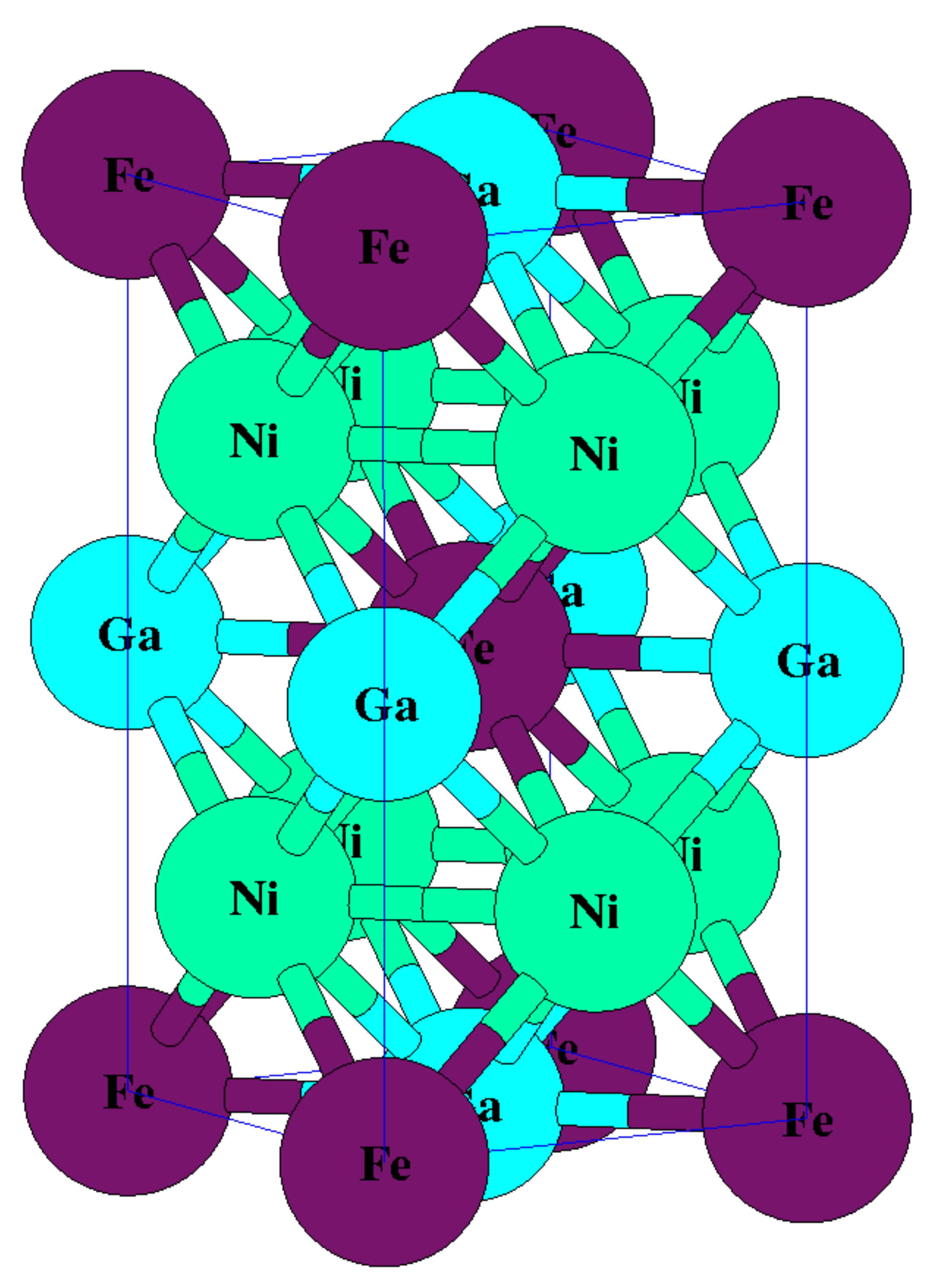}}

  \caption{(Color online)Structures of (a) Austenite (b) Martensite phases of Ni$_2$FeGa}
\end{figure}
 
\begin{figure}[h]

  \centering
   \includegraphics[scale=0.27]{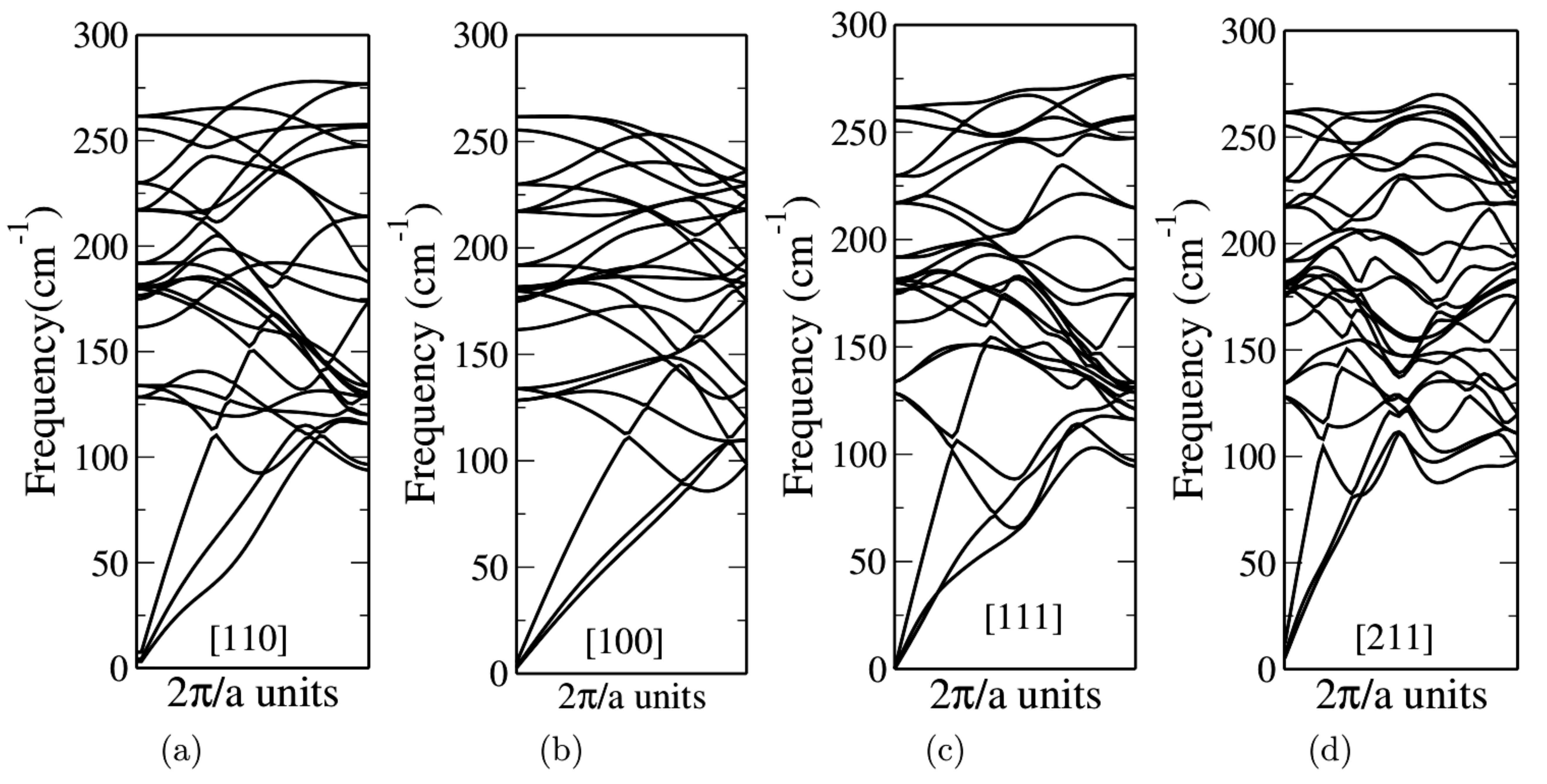}
  \caption{Phonon spectra of  martensite Ni$_2$FeGa along different high symmetry directions: (a) [110] direction (b) [100] direction (c) [111] direction (d) [211] direction}
\end{figure}
\begin{figure}[h]
  \centering
  \includegraphics[scale=0.27]{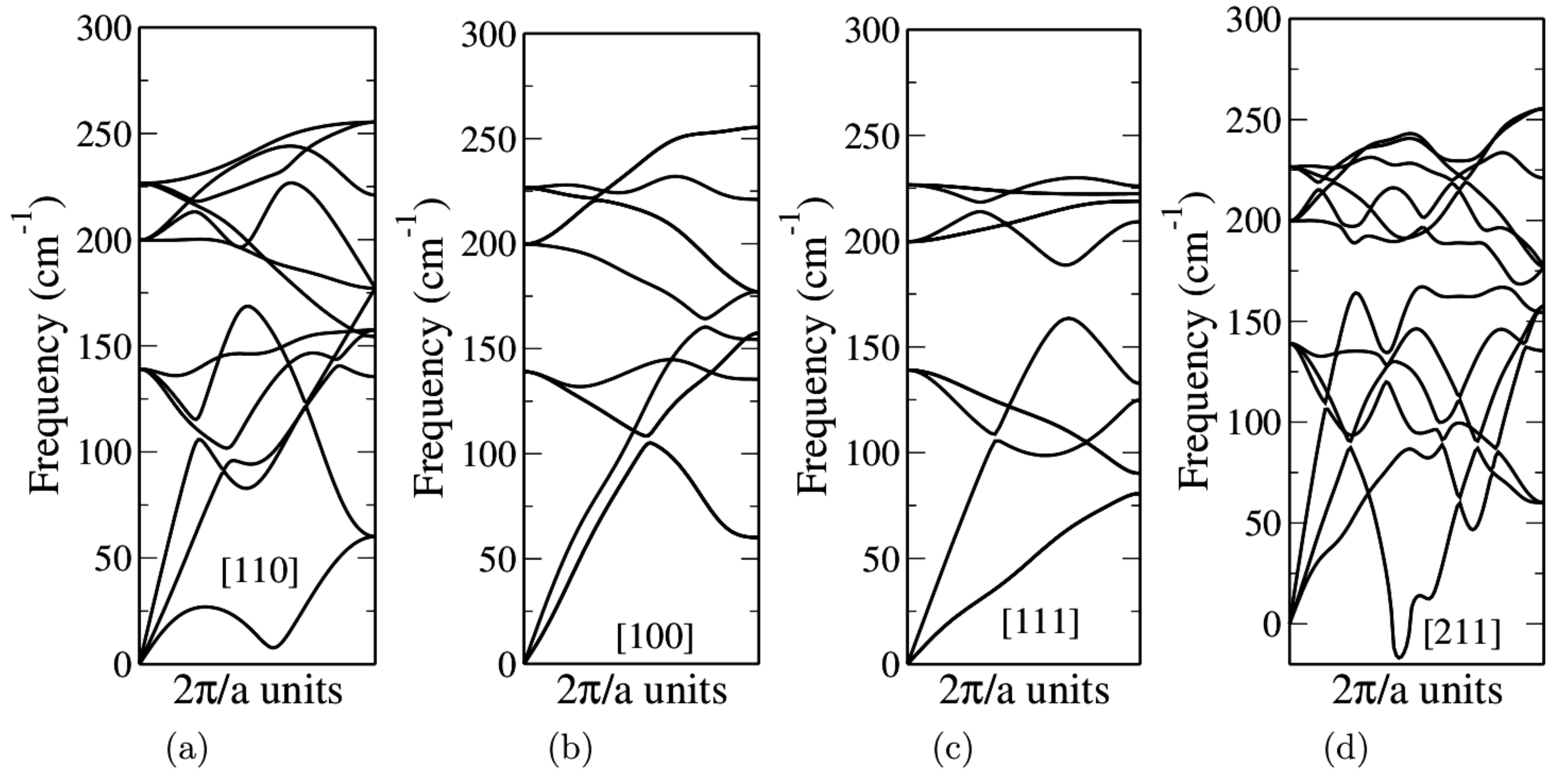}
     
  \caption{Phonon spectra of austenite Ni$_2$FeGa  along different high symmetry directions: (a) [110] direction (b) [100] direction (c) [111] direction (d) [211] direction}
\end{figure}  
\begin{figure}[h]
  \centering
  \includegraphics[scale=0.3]{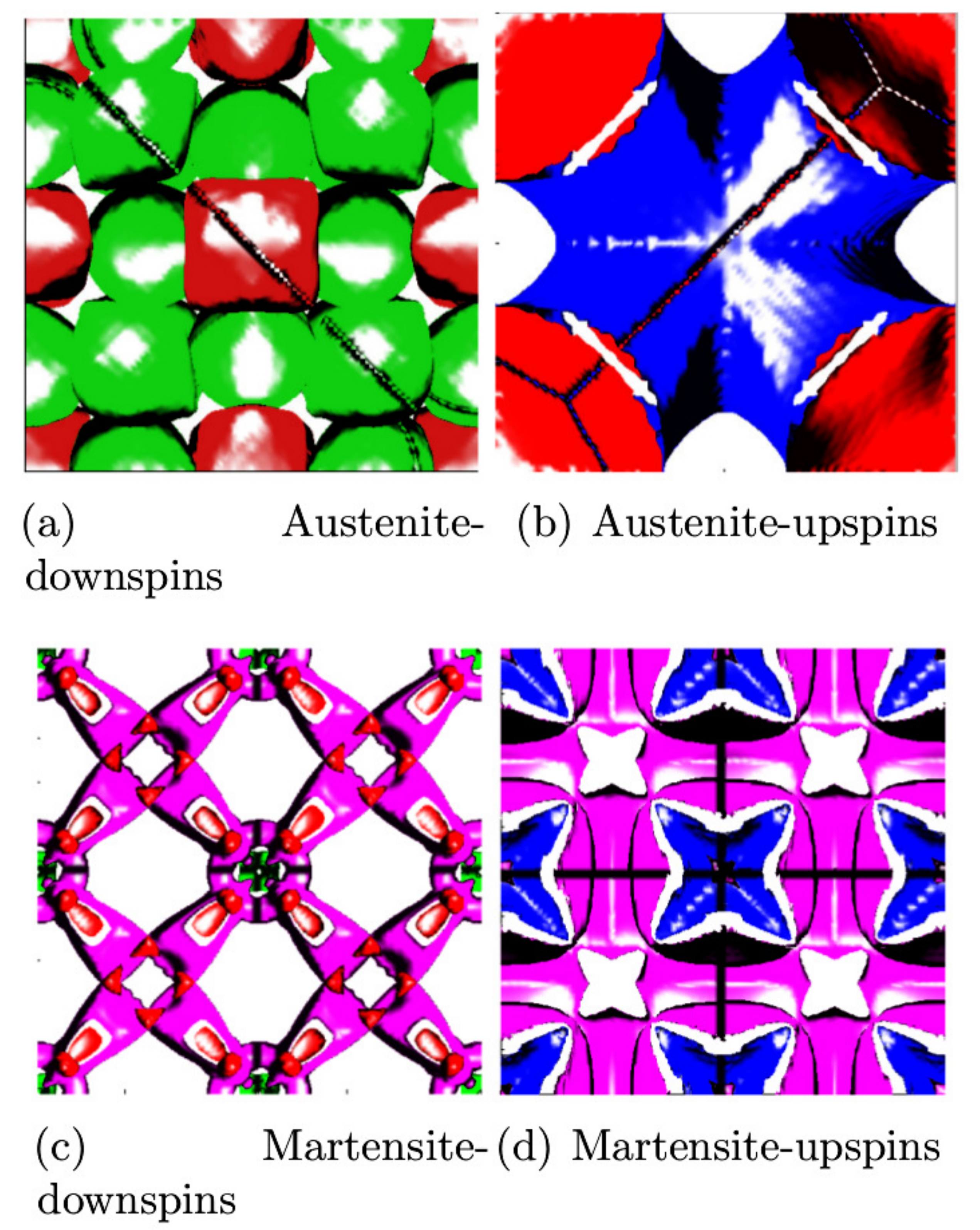}
  \caption{(color online)3D plot of Fermi surface of Ni$_2$FeGa in both austenite and martensie phase (a) 19th(green) and 20th(red) austenite downspin bands (b) 22nd(red) and 23rd(blue) austenite upspin bands (c) 37th(off-red), 38th(red) and 39th(green) martensite downspin bands (d) 44th(blue) and 45th(off-red) martensite upspin bands}
\end{figure}  
Optimized structural parameters from our previous work on Ni$_2$FeGa have been used for both austenite and martensite phase calculations\cite{Sah}. The cubic structure of austenite phase (L2$_1$) and the tetragonal structure of martensite phase (L1$_0$) are shown in Fig. 1.  Figures 2(a), 2(b), 2(c), 2(d) show the phonon spectra calculated along different high symmetry directions [110], [100], [111], [211] for the martensite phase. The low lying acoustic phonon modes are less populated as compared to the higher optical phonon modes. From the figures, it is observed that the phonon branches in the martensite phase show neither imaginary frequencies nor significant phonon dip in any of the different high symmetry directions. The calculated phonon spectra along different high symmetry directions [110], [100], [111], [211] for the austenite phase are shown in Fig. 3(a), 3(b), 3(c), 3(d). Here the low lying acoustic phonons show different behaviour. The transverse acoustic phonon mode TA$_2$ shows softening in [110] and [211] directions. It has also been observed in our previous work that the TA$_2$ phonon branch along [110] direction shows softening at a wave-vector $\textbf{q}=0.58[110]$ and this anomaly has been related to the Fermi surface nesting\cite{Sat}. This type of phonon anomaly along [110] direction has also been related with generalised susceptibility peaks and Fermi nesting vectors in other similar alloys\cite{Cla,Vel}. The other less explored anomaly that we observed for the first time is the complete softening of the TA$_2$ phonon branch along [211] direction in austenite Ni$_2$FeGa system. This is in exact correpondence to the Electron Diffraction results of J. Q. Li \textit{et. al.}\cite{Li} where they found a new structural anomaly along [211] direction.  Their finding reveals diffused satellite peaks along [211] and [110] directions that correspond to micromodulated structures in Ni$_2$FeGa system. From Fig. 3(d), it is observed that there are two dips in TA$_2$ branch along [211] direction. The first big dip at $\textbf{q}=0.4$ gets completely softened going well down to negative frequencies. The second dip at $\textbf{q}=0.6$ is very close to the phonon anomaly along [110] direction. It can be inferred that the first dip  at $\textbf{q}=0.4$ becomes more significant at the expense of the second dip at $\textbf{q}=0.6$. This is indicative of the fact that the phonon anomaly wavevector oscillates between $\textbf{q}=0.6$ in [110] plane and $\textbf{q}=0.4$ in [211] plane. This point is in good agreement with the experimental findings of J. Q. Li \textit{et. al.} where the two structural modulations along [110] and [211] directions altered systematically with low temperature phase transition.  So, from lattice dynamical point of view, the martensite phase is more stable than the austenite phase. The instability which is shown in the austenite phase leads to a metastable phase where different structural modulations occur.

\begin{figure}[htp] 
\centering
  \includegraphics[scale=0.27]{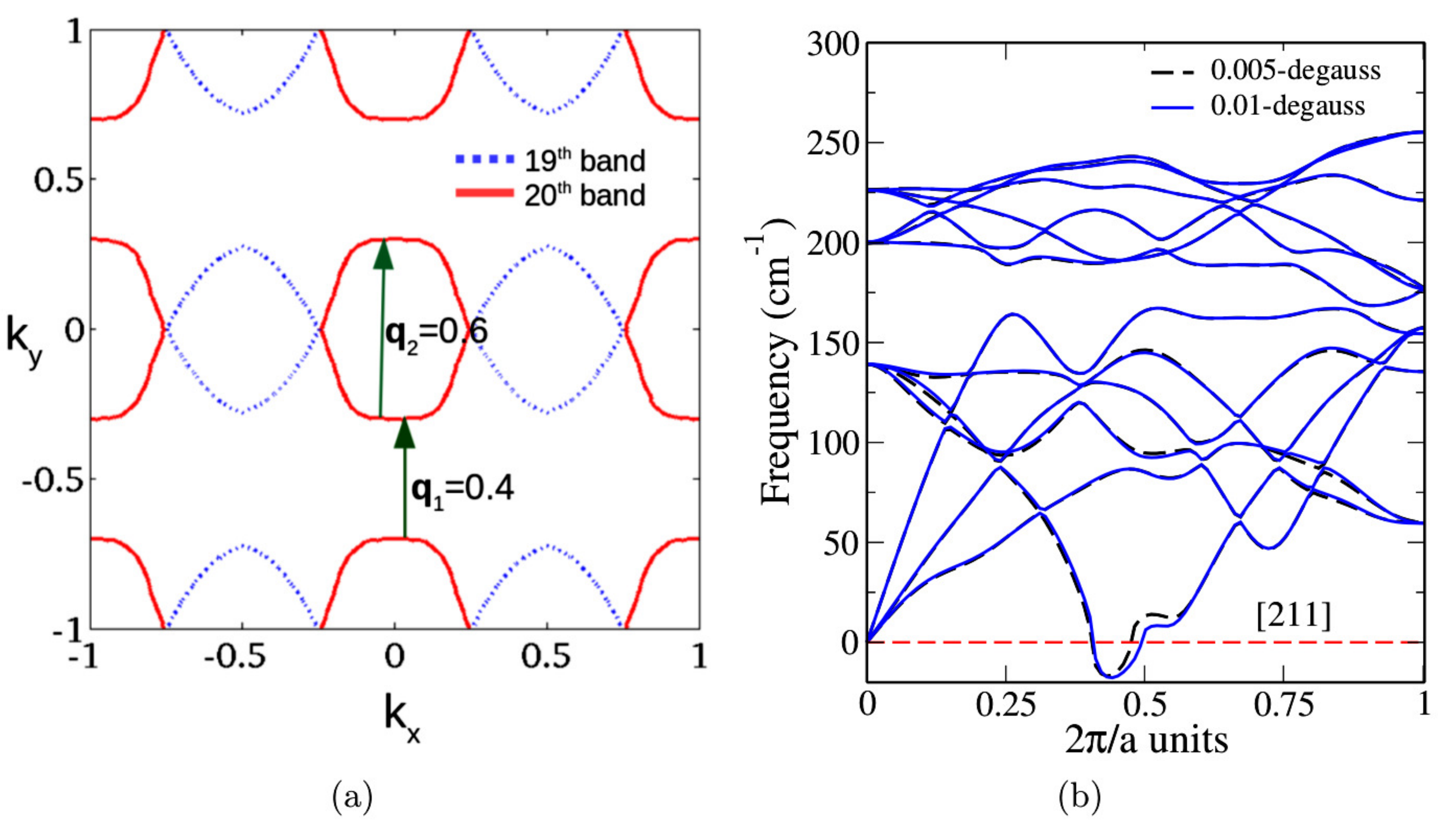}
  \caption{(color online)(a) 2D Fermi surface cross sections of the 19-20th downspin bands along (211) plane in austenite phase (b) Phonon curve along [211] direction for austenite phase Ni$_2$FeGa at degauss values 0.005 and 0.01 Rydberg}
\end{figure}
\begin{figure}[htp]
  \centering
  \subfigure[]{\includegraphics[scale=0.35]{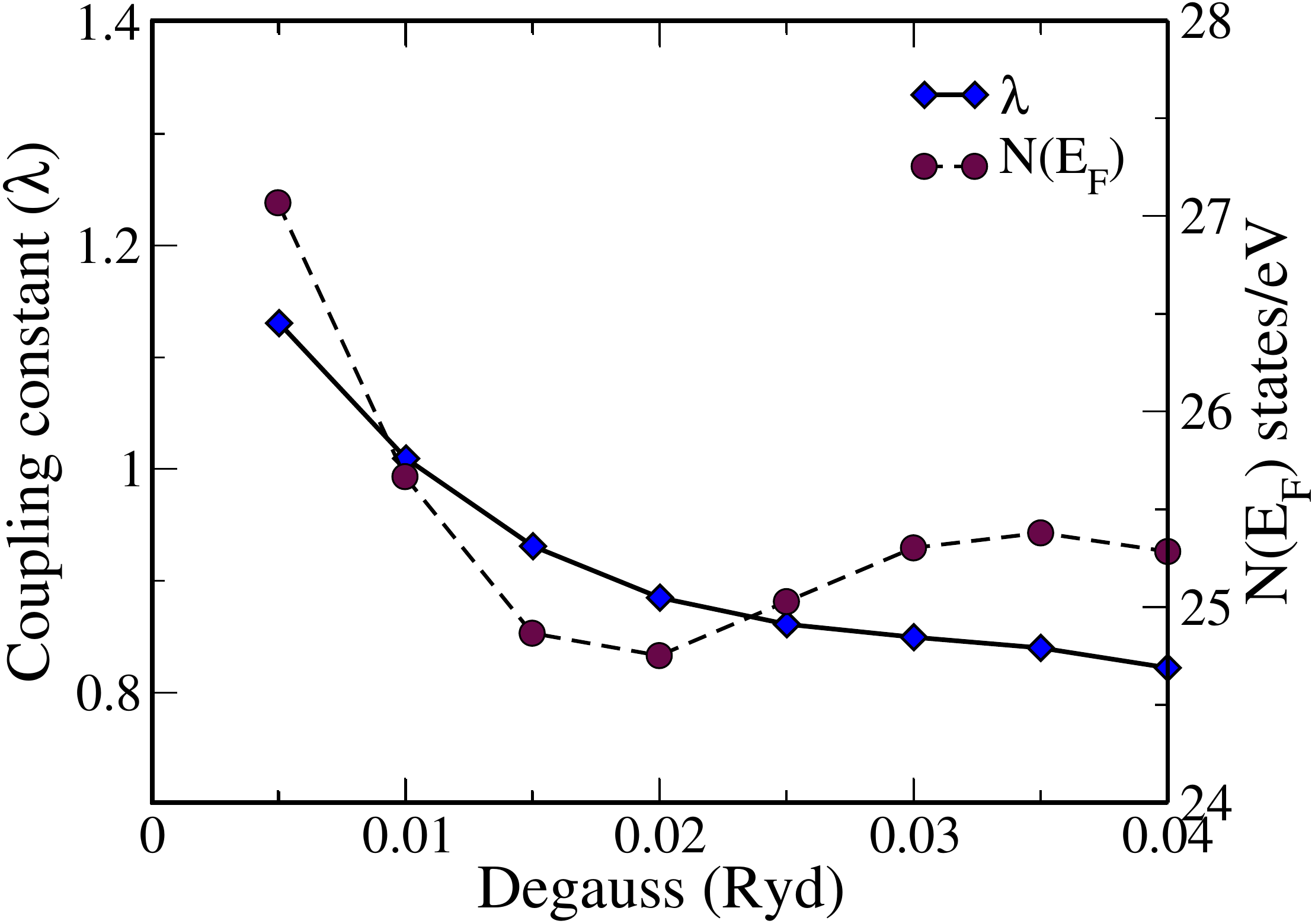}}
  \subfigure[]{\includegraphics[scale=0.33]{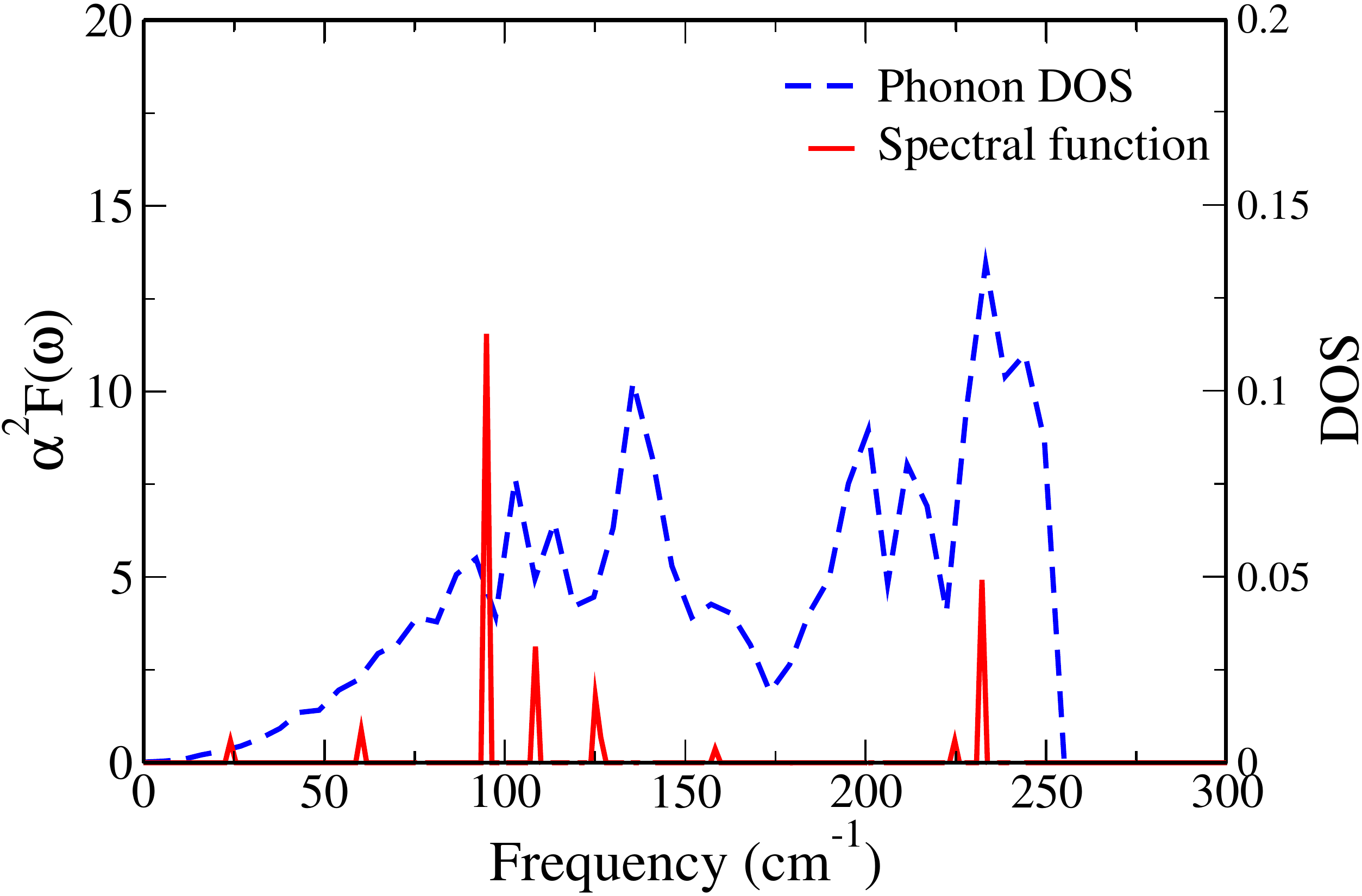}}   
  
  \caption{(color online)Austenite phase: (a) Electron phonon coupling constant($\lambda$ ) and DOS at the Fermi level N(E$_F$) versus degauss values (b) Spectral function and Phonon DOS versus phonon frequency }
\end{figure}
\bigskip

The 3D Fermi surface of both austenite and martensite phases have been calculated and the plots are shown in Fig.4. The upspin and downspin bands have been seperated for both the phases. Figure 4(a) shows the 19th(green) and 20th(red) downspin bands of the austenite phase. Figure 4(b) shows the 22nd(red) and 23rd(blue) upspin bands  of the austenite phase. Figure 4(c) shows the 37th(off-red), 38th(red) and 39th(green)  downspin bands  of the martensite phase. Figure 4(d) shows the 44th(blue) and 45th(off-red)  upspin bands of the martensite phase. From Fig. 4(a) and 4(b), it is seen that in the austenite phase, the collective Fermi surface is dense for downspin electrons having distinctly seperate surfaces for each band type while the upspin bands have surfaces which are continuous in nature.  In austenite-downspin bands, there are several parallel flat surfaces for each type of band increasing the nesting feature which is not observed in upspin bands. In case of the martensite phase, the observed 3D Fermi surface in both upspins and downspin bands are not so dense as compared to the above case of austenite phase as shown in Fig. 4(c) and 4(d). As a cosequence, Fermi surfaces for both spin types in the martensite phase do not show any extended parallel flat surfaces that could give rise to nesting in this phase. So, no Fermi surface nesting is observed in the martensite phase of this alloy. This can be expected as we did not see any phonon anomaly in our above calculations on phonon spectrum in the martensite phase. Moreover, this point is further supported by our previous calculations on electronic structure where we found the density of states(DOS) in the austenite phase to be much larger than the DOS in the martensite phase at the Fermi level\cite{Sah}. As it was observed in our previous calculations that the downspin electrons are responsible for destabilizing the austenite phase, we cut the 3D Fermi surface of the austenite-downspin bands to get the 2D cross sections. As the dynamical instability in the [110] direction has already been addressed in the previous work\cite{Sat}, we focus only on the second type of instability observed along [211] direction.  Figure 5(a) shows the 2D cross sections of the 19th and 20th austenite-downspin bands in (211) plane. Here we can clearly see intraband nesting features in the Fermi surface of the 20th band. Two different nesting vectors are observed with magnitudes $\textbf{q}_1=0.4$ and $\textbf{q}_2=0.6$. These two wavevectors exactly match the phonon anomaly wavevectors along [211] direction that we observed in our above phonon calculations. Figure 5(b) shows the phonon spectra along [211] direction for two different degauss values 0.005 Rydberg and 0.01 Rydberg. It is observed that the two curves coincide almost in all modes except the TA$_2$ branch which shows a slight difference. Qualitatively, the two TA$_2$ branches show the same behaviour without any change in the position of wavevectors of the phonon dips. The q-mesh that we used here for lattice dynamical calculations is $6\times6\times6$ Monkhorst Pack. This value is good enough to come over the numerical instabilities and produce a converged result.

Above calculations show that the phonon anomalies observed from lattice dynamical calculations can be correlated well with the Fermi nesting vectors observed from electronic structure calculations in this system. It can be expected that there will be significant coupling of electrons and phonons as these two parameters are well intertwined in this type of system. To probe this idea, calculations on electron-phonon interaction has been carried out using Density Functional Perturbation Theory(DFPT). The parameter which governs electron-phonon coupling(EPC) is the Eliashberg function $\alpha^2 F(\omega,\epsilon,k)$ which is equal to the transition probability from and to a photohole state $(\epsilon,k)$ via
coupling to a phonon mode $\omega$. The  total EPC constant $\lambda$ is given by,
$$ \lambda= 2\int^\infty_0 \dfrac{\alpha^2 F(\omega)}{\omega}d\omega $$
Eliasberg's function has been calculated to get the electron-phonon coupling(EPC) coefficients.  
Figure 6(a) shows the variation of electron-phonon coupling constant $\lambda$  and DOS at the Fermi level (E$_F$) with respect to Gaussian broadening i.e. degauss parameter. The basic idea behind Gaussian broadening is that it just takes care of the partial occupancies that come up in this metallic system. The DOS at the Fermi level becomes minimum at a degauss value of 0.02 Rydberg.  It is also seen that the convergence of the coupling constant $\lambda$ occurs around 0.02 Rydberg which gives $\lambda$= 0.90. This shows that there is significant coupling of the electrons and phonons in the system. Figure 6(b) shows the Eliashberg function or the spectral function $\alpha^2 F(\omega)$ plot at a degauss value of 0.02 Rydberg and phonon density of states with respect to frequency.  From the peaks of the spectral function, it is observed that the electron-phonon coupling is mainly contributed from four different frequencies. The first main contribution comes from 95 $cm^{-1}$ and other significant contributions from 110 $cm^{-1}$, 127 $cm^{-1}$ and 233 $cm^{-1}$. It may be pointed out that the maximum electron-phonon coupling occurs at 95 $cm^{-1}$ which is a transition frequency region from lower acoustic modes to higher optical modes. So there is a strong belief that the electron-phonon coupling in this system is also mediated through lattice coupling between lower frequency acoustic modes and higher frequency optical modes.
\bigskip
\section{CONCLUSIONS}
Our calculation supports the presence of phonon anomaly along [211] direction  that was observed experimentally for the first time by J. Q. Li \textit{et. al.} as structural modulations along [211] direction in Ni$_2$FeGa system. 2D and 3D Fermi surface calculations in the austenite phase have shown Fermi nesting features in (211) plane with magnitude of wavevectors exactly matching the wavevectors of phonon anomaly along [211] direction. This infers that the observed Fermi nesting vectors in the system might be responsible for the observed lattice instabilities along [211] direction. Electron-phonon calculation in the austenite phase shows that there is significant  coupling of electrons and phonons in the system and maximum contribution comes from the transition frequency region from lower acoustic modes to higher optical modes. From all the above calculations, it is clear that the phonon anomalies, the Fermi surface nesting features and electron-phonon coupling parameters in this system are closely related to one another. Finally, we are able to give a proper theoretical explanation of the various microscopic parameters that drives the technologically important martensitic transformation phenomena in Ni$_2$FeGa system to some extent. This type of theoretical investigations on shape memory alloys might be useful to get an understanding on how to design and manufacture appliances and devices using SMAs for day-to-day and industrial applications.
\bigskip

\section{ACKNOWLEDGEMENTS}
 MBS would like to acknowledge the support from DST, Government of India, through Project No. SR/FTP/PS-031/2012
 

\end{document}